\documentstyle[12pt,html,epsf,psfig]{article}

\begin{document}
\title{MATTERS OF GRAVITY, The newsletter of the APS Topical Group on 
Gravitation}
\begin{center}
{ \Large {\bf MATTERS OF GRAVITY}}\\ 
\bigskip
\hrule
\medskip
{The newsletter of the Topical Group on Gravitation of the American Physical 
Society}\\
\medskip
{\bf Number 22 \hfill Fall 2003}
\end{center}
\begin{flushleft}

\tableofcontents
\vfill
\section*{\noindent  Editor\hfill}

Jorge Pullin\\
\smallskip
Department of Physics and Astronomy\\
Louisiana State University\\
Baton Rouge, LA 70803-4001\\
Phone/Fax: (225)578-0464\\
Internet: 
\htmladdnormallink{\protect {\tt{pullin@phys.lsu.edu}}}
{mailto:pullin@phys.lsu.edu}\\
WWW: \htmladdnormallink{\protect {\tt{http://www.phys.lsu.edu/faculty/pullin}}}
{http://www.phys.lsu.edu/faculty/pullin}\\
\hfill ISSN: 1527-3431
\begin{rawhtml}
<P>
<BR><HR><P>
\end{rawhtml}
\end{flushleft}
\pagebreak
\section*{Editorial}

Not much to report here. Just to remind people that the year 2005
is the ``year of physics'' commemorating Einstein's remarkable year
of 1905. The Topical Group has set up a committee chaired by 
Richard Price to coordinate activities. There is a worldwide site
\htmladdnormallink{http://www.physics2005.org}
{http://www.physics2005.org}.
I want to encourage the readership to suggest
topics for articles in MOG. In the last few issues articles were
solicited by myself. This is not good for keeping the newsletter
balanced. Either contact the relevant correspondent or me directly.

The next newsletter is due February 1st. All issues are available in
the WWW:\\\htmladdnormallink{\protect
{\tt{http://www.phys.lsu.edu/mog}}} {http://www.phys.lsu.edu/mog}\\

The newsletter is  available for
Palm Pilots, Palm PC's and web-enabled cell phones as an
Avantgo channel. Check out 
\htmladdnormallink{\protect {\tt{http://www.avantgo.com}}}
{http://www.avantgo.com} under technology$\rightarrow$science.

A hardcopy of the newsletter is distributed free of charge to the
members of the APS Topical Group on Gravitation upon request (the
default distribution form is via the web) to the secretary of the
Topical Group.  It is considered a lack of etiquette to ask me to mail
you hard copies of the newsletter unless you have exhausted all your
resources to get your copy otherwise.  

If you have comments/questions/complaints about the newsletter email
me. Have fun.

\bigbreak
   
\hfill Jorge Pullin\vspace{-0.8cm}
\parskip=0pt
\section*{Correspondents of Matters of Gravity}
\begin{itemize}
\setlength{\itemsep}{-5pt}
\setlength{\parsep}{0pt}
\item John Friedman and Kip Thorne: Relativistic Astrophysics,
\item Raymond Laflamme: Quantum Cosmology and Related Topics
\item Gary Horowitz: Interface with Mathematical High Energy Physics and
String Theory
\item Beverly Berger: News from NSF
\item Richard Matzner: Numerical Relativity
\item Abhay Ashtekar and Ted Newman: Mathematical Relativity
\item Bernie Schutz: News From Europe
\item Lee Smolin: Quantum Gravity
\item Cliff Will: Confrontation of Theory with Experiment
\item Peter Bender: Space Experiments
\item Riley Newman: Laboratory Experiments
\item Warren Johnson: Resonant Mass Gravitational Wave Detectors
\item Stan Whitcomb: LIGO Project
\item Peter Saulson: former editor, correspondent at large.
\end{itemize}
\section*{Topical Group in Gravitation (GGR) Authorities}
Chair: John Friedman; Chair-Elect: Jim Isenberg; Vice-Chair: 
Jorge Pullin;
Secretary-Treasurer: Patrick Brady; Past Chair: Richard Price; 
Members at Large:
Bernd Bruegmann, Don Marolf, 
Gary Horowitz, Eric Adelberger, 
Ted Jacobson, Jennie Traschen. 
\parskip=10pt
\vfill
\pagebreak

\section*{\centerline {
We hear that...}}
\addtocontents{toc}{\protect\medskip}
\addtocontents{toc}{\bf GGR News:}
\addcontentsline{toc}{subsubsection}{\it  
We hear that... by Jorge Pullin}
\begin{center}
Jorge Pullin, Louisiana State University
\htmladdnormallink{pullin@phys.lsu.edu}
{mailto:pullin@phys.lsu.edu}
\end{center}

{\em Saul Teukolsky} was elected to the National Academy
of sciences.

{\em Bill Unruh} was elected to the American Academy of 
Arts and Sciences.

{\em Yvonne Choquet-Bruhat and Jimmy York} were awarded the
Dannie Heinemann prize of the American Physical Society.

Hearty congratulations!
\vfill
\eject

\section*{\centerline {
Update on a busy year for LIGO}}
\addtocontents{toc}{\protect\medskip} \addtocontents{toc}{\bf
Research Briefs:} \addcontentsline{toc}{subsubsection}{\it Update
on a busy year for LIGO, by Stan Whitcomb}
\begin{center}
Stan Whitcomb, LIGO-Caltech, on behalf of the LIGO Scientific
Collaboration \htmladdnormallink{whitcomb\_s@ligo.caltech.edu}
{mailto:whitcomb\_s@ligo.caltech.edu}
\end{center}

2003 marks another important year for LIGO as it continues the
transition from construction and commissioning to full science
data-taking.  The first papers describing searches for
gravitational waves using data from the first Science Run between
the LIGO and GEO detectors are being submitted, a second Science
Run has been accomplished with an order of magnitude improvement
in sensitivity, commissioning activities to bring the detectors to
their design performance are making good progress, and planning
for an even more sensitive set of detectors (``Advanced LIGO'') is
well underway.

{\bf Results from LIGO's first Science Run}

The first LIGO science run [1], S1, spanned a period of 17 days in
August and September 2002; the GEO-600 interferometer also
operated during S1.  The three LIGO detectors were operating with
a noise level about a factor of 100 above the design level, so the
probability of detecting any gravitational waves, particularly in
such a short run were extremely small.  However, the upper limits
which could be set from these observations on different types of
gravitational waves are typically as good or better than previous
direct observational limits.

The LIGO Scientific Collaboration has taken the leadership in
analyzing the S1 data, forming four data analysis working groups,
each aimed at a particular type of gravitational wave signal. One
working group focused on searching for gravitational waves from
the inspiral of binary neutron stars.  A second one searched for
the nearly sinusoidal waves from a millisecond pulsar
(J1939+2134).  The third one searched for a stochastic background
of gravitational waves, while the fourth looked for poorly
modeled burst-type sources such as supernovae or GRBs.

In the previous issue of MOG [2], Gary Sanders gave a description
of preliminary results from S1, using both LIGO and GEO-600 data.
These results have been refined, and four papers, one for each
analysis, are in the final stages of preparation prior to
submission. The performance parameters of the LIGO and GEO-600
detectors during S1, including their configurations, plots of
sensitivity, and tables giving the fraction of time that each
detector was operational, are given in a fifth paper.  Preprints
of these papers are being made available [links at reference 3] as
they are approved by the LIGO Scientific Collaboration.  The
publication of these papers marks a real milestone in LIGO's
evolution.

{\bf The second Science Run}

S1 was the first in a series of progressively more sensitive
science runs, interleaved with interferometer commissioning. The
second Science Run, S2, took place from February 14 to April 14,
2003. The interferometers showed good reliability for this stage
in their development.  The duty cycle for the interferometers,
defined as the fraction of the total run time when the
interferometer was locked and in its low noise configuration,
ranged between 37\% (for the Livingston interferometer) and 74\%
(for the Hanford 4 km interferometer).  Building on the lessons
learned from S1, procedures were put in place to better monitor
the performance and the result was better stability of the
performance.

Most important, S2 sensitivity was improved by more than an order
of magnitude over S1. Typical noise levels for S2 were about a
factor of 10 better than in S1 with the 4km interferometer at
Livingston having the greatest sensitivity, followed by the 4 km
interferometer at Hanford.  The increase sensitivity represents
the first time the LIGO detectors have had the sensitivity to see
potential sources in other galaxies, most notably the Andromeda
galaxy. Once again, LIGO would not be expected to see
gravitational waves at this level, but it still represents a
significant step toward design sensitivity.

The S2 run also involved coincident running with the TAMA-300
detector, following the signing of a new LIGO-TAMA Memorandum of
Understanding.

{\bf Commissioning progress}

Since the end of S2, the commissioning team has been hard at work
to complete an ambitious set of improvements and ``fixes'' to the
interferometers at both sites.

At both sites, there have been changes inside the vacuum system.
Minor changes to the positions of optics have been made to made to
adjust cavity lengths closer to their design values.  At Hanford,
the input test mass on one arm was replaced because of a lossy AR
coating.  Baffles have been installed to prevent errant high power
laser beams from damaging suspension wires on the input optics.

A number of changes have been made to reduce the coupling of
acoustic noise to the interferometers.  Steps have been taken to
reduce the amount of acoustic noise generated by the air
conditioning system and by fans in the electronics racks. Acoustic
enclosures have been purchased to surround the most sensitive
optical tables.  Larger aperture optics and mounts have been
installed on the optical tables at the outputs of the
interferometer.

The commissioning of the wavefront-sensing alignment systems have
been a high priority at both sites.  This system was operating to
control 8 of 10 angular degrees of freedom on the 4km
interferometer at Hanford during S2, and demonstrated how much
more stability could be achieved.  Good progress has been made on
both all interferometers.

As a result of the various changes, the 4 km interferometer at
Hanford has achieved its highest sensitivity yet, a range for
binary neutron star inspirals of 1.5 Mpc.

{\bf Future plans}

Two major activities loom on the near-term horizon for LIGO: the
third Science Run (S3) and an upgrade to the Seismic Isolation
System at the LIGO Livingston Observatory (LLO).

S3 is scheduled for approximately two months in November and
December 2003.  The commissioning progress described in the
previous section should provide a significant sensitivity
improvement over S2, though perhaps not as great as the jump
between S1 and S2.  The more complete implementation of the
wavefront-sensing alignment system should give better stability
and move the analysis closer to what is expected in full
operation.

Immediately after S3, we plan to install an upgrade to the Seismic
Isolation System at LLO. Since the beginning of commissioning, the
interferometer at LLO has suffered from higher than expected
seismic noise due to anthropogenic sources.  These large motions
occur at low frequencies and often exceeded the ability of the
control systems to cope with them.  As a result, the duty cycle of
the LLO interferometer has been lower than specified. The upgrade
consists of active vibration isolation systems installed outside
the vacuum system to cope with ground motions in the 0.1-10 Hz
range.  It has been under development for the past 18 months,
initially aimed at Advanced LIGO, but once it was determined that
it could help the situation at LLO, its development was
accelerated. Hardware is expected to be ready for installation in
very early 2004.

{\bf Advanced LIGO}

The LIGO Laboratory, with strong support from the LIGO Scientific
Collaboration, submitted its proposal to the NSF in March, 2003
for Advanced LIGO. The proposed system consists of three nominally
identical interferometers -- two 4km systems at Hanford, and one
at Livingston, and each are tunable through variations of the
input power and the signal recycling mirror position. The increase
in sensitivity is greater than a factor of 10 over initial LIGO,
and also the potential for observation a factor of 4 lower in
frequency. The result is that it is anticipated that one will be
able to see e.g., 1.4 solar mass neutron star binary inspiral
signals to roughly 350 Mpc (for the 3 interferometer detector
system), or greater than a factor of 15 further than initial LIGO
for this source. This new detector, to be installed at the LIGO
Observatories, will replace the present detector once it has
reached its goal of a year of observation, with the planning date
for first observations presently at 2010. The improvement of
sensitivity will allow the one-year planned observation time of
initial LIGO to be equaled in just several hours.

The new design involves a complete replacement of the initial
detector. The laser power is increased from 10 W to 180 W, to
improve the shot-noise limited sensitivity of the instrument --
this will be a contribution from the Hannover GEO colleagues.
Prototypes have demonstrated over 100 W to date. The input optics,
under the leadership of the University of Florida LIGO group, will
resemble the initial LIGO design but improved to handle the higher
power. The test masses will be made of sapphire, 40 kg in mass to
resist the photon pressure fluctuations, and with coatings which
must be low optical loss and low mechanical loss (to keep thermal
noise low). Sapphire 'pathfinder' pieces of the final size have
been fabricated, and show excellent mechanical properties. The
test mass suspensions, contributed by our Glasgow GEO colleagues,
resemble the GEO 600 design with a final stage using fused silica
fibers, again for low thermal noise. The seismic isolation design
comes from LSU and Stanford, and uses high-gain servo systems to
deliver very low motion in as well as below the gravitational wave
band -- the low-frequency noise of the interferometer at low power
will be limited by the Newtonian background from gravitational
gradients. Aspects of the design have been tested in various
prototypes, and a complete prototype is in test at Stanford at
this time. These mechanical aspects of the design will be tested
together at the MIT LASTI full-scale test facility. The
gravitational readout system will use a form of DC sensing, moving
slightly away from the dark fringe of the interferometer output
port, and will make use of the innovations from table-top and
suspended signal-recycled interferometer tests made in Australia,
Florida, Glasgow, Garching, and Caltech; a complete engineering
test of the readout and control system is in development at the
40m Lab at Caltech.

A review of the proposal was held by the NSF in June, and the
feedback was quite positive. The materials for the review can be
found at \htmladdnormallink{http://www.ligo.caltech.edu/advLIGO/}
{http://www.ligo.caltech.edu/advLIGO/}, and can serve as a
resource for further information. We are excited by the kind of
leap forward this instrument should give to the field, and hope
that it will be observing in concert with other instruments that
can be operating at that time -- a second generation VIRGO, and a
cryogenically-cooled underground system in Japan as examples. When
will it be another ordinary day when a few more gravitational
waves sources are logged? We hope around 2010!

LIGO is funded by the US National Science Foundation under
Cooperative Agreement PHY-0107417. This work is a collaborative
effort of the Laser Interferometer Gravitational-wave Observatory
and the institutions of the LIGO Scientific Collaboration. LIGO
T030185-00-D.

More information about LIGO can be found at:
\htmladdnormallink{http://www.ligo.caltech.edu}
{http://www.ligo.caltech.edu}.

{\bf References:}

[1] MOG article describing S1 performance: 

\htmladdnormallink
{http://www.phys.lsu.edu/mog/mog20/node10.html}
{http://www.phys.lsu.edu/mog/mog20/node10.html}

[2] MOG article describing S1 results: 

\htmladdnormallink
{http://www.phys.lsu.edu/mog/mog21/node10.html}
{http://www.phys.lsu.edu/mog/mog21/node10.html}

[3] S1 paper links: 

\htmladdnormallink
{http://www.ligo.caltech.edu/LIGO\_web/s1/}
{http://www.ligo.caltech.edu/LIGO\_web/s1/}

\vfill \eject

\section*{\centerline {First year results from the}\\
\centerline{Wilkinson Microwave Anisotropy Probe (WMAP)}}
\addcontentsline{toc}{subsubsection}{\it 
First Year Results From WMAP, by Rachel Bean}
\begin{center}
Rachel Bean, Princeton
Collaboration \htmladdnormallink{rbean@astro.princeton.edu}
{mailto:rbean@astro.princeton.edu}
\end{center}

The cosmic microwave background (CMB), along with the distribution of
large scale structure, has become one of the principal tools for
deciphering the cosmological content and history of the universe. The
WMAP satellite, launched on June 30, 2001, completed its first full
year of measurements of the CMB in August 2002, with the data analyzed
and published earlier this year. This article presents a brief summary
of the approach and key findings from the mission's first results. For
further details see [1] and the 12 companion papers
referred to within it, in particular this article focuses on the
cosmological parameters extracted from the data discussed in
[2].

For those not familiar with the CMB, it is comprised of photons that
interacted strongly with the plasma of free electrons and baryonic
ions in the early universe. At this time the photon mean free path was
short and the universe was effectively opaque. As the universe
expanded and subsequently cooled below 3000K, 380,000 years after the
Big Bang, electrons recombined with nuclei and fewer charged particles
were present to interact with the CMB photons, the photons `decoupled'
from the rest of the matter and the universe became transparent. The
distribution of temperature and polarization fluctuations that we
measure today in the CMB are therefore effectively those imprinted at
the epoch of recombination,``the decoupling surface".

Within the last decade, starting with the results from COBE
[3], a plethora of experiments have measured the anisotropy in
the fluctuations in the CMB temperature. Together they have
incontrovertibly detected the first acoustic peak of oscillations in
the CMB power spectrum. This peak arises from oscillations in the
coupled photon- baryon fluid just prior to when photons decoupled and
is direct experimental support for the CMB being decoupled photons and
for the standard recombination model. In addition, last year, there
was the first detection of anisotropy in the polarization of the CMB
[4].

WMAP was created with the aim of extending on previous observations in
two main ways: to make a map of the full sky, and to measure the CMB
with much improved precision by minimizing systematic errors. The
precision is obtained through measuring the CMB over five frequency
bands, which allow external contaminants such as dust and point
sources to be removed more efficiently. WMAP observes the sky
convolved with the beam pattern (the ``window function") of the
detectors. Imperfect knowledge of the window function is one of the
main internal systematics and therefore minimizing this uncertainty by
accurate in-flight determination of the beam patterns has also been a
key factor in achieving WMAP's precision. Figure 1 shows the improved
resolution of the WMAP results in comparison to the only previous full
sky map, that of COBE.  Also shown is the power spectrum of
fluctuations measured by WMAP for temperature-temperature ``TT" and
temperature-polarization ``TE" correlations , in multipoles, $l$, from
spherical harmonic decomposition of the sky,.

\begin{figure}[t]
\centerline{
\psfig{file=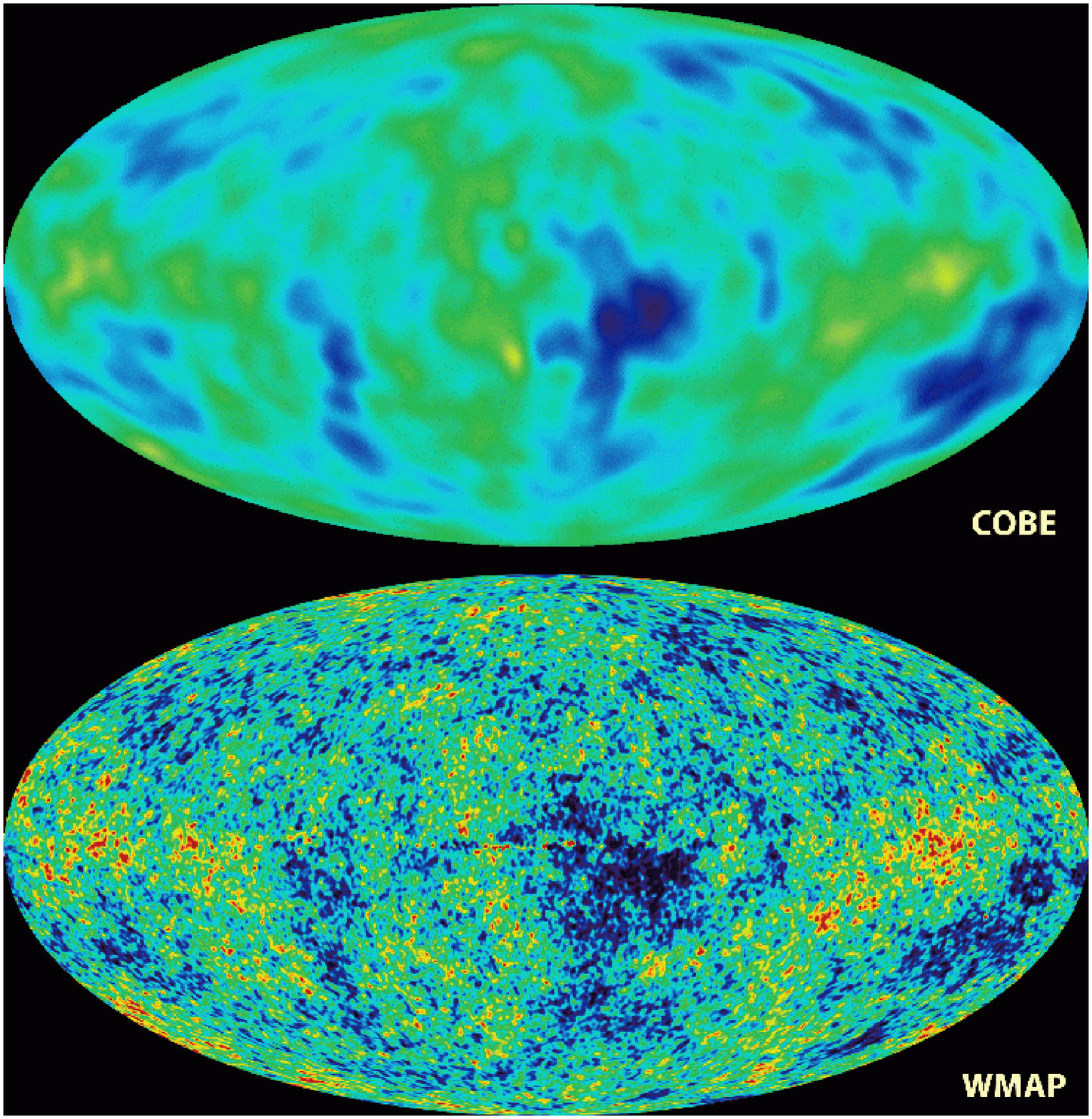,width=2.5in,height=2.75in} 
\psfig{file=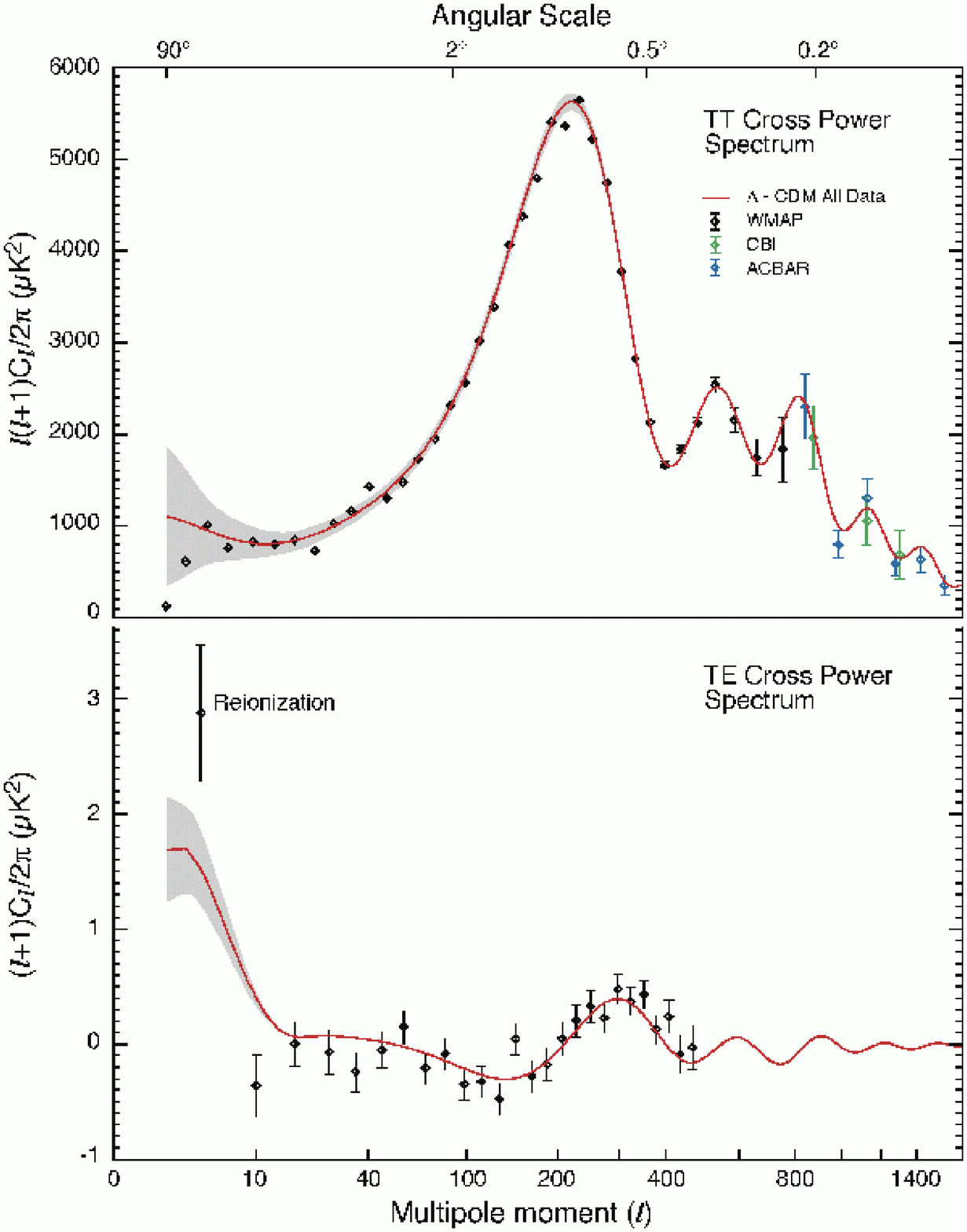,width=2.5in,height=2.75in}}
\caption{Left panel: An all-sky image of the Universe 380,000 years
after the Big Bang. In 1992, NASA's COBE mission first detected tiny
temperature fluctuations (shown as color variations) in the infant
universe. The WMAP's improved resolution brings the COBE picture into
sharper focus. Right panel: The ``angular spectrum" of the fluctuations
in the WMAP full-sky map. The top curve shows the power spectrum for
the temperature fluctuations, while the lower curve shows the cross
correlation of temperature with polarization. In each figure the best
fit cosmological model is shown in red for the `standard' scenario
discussed in the text. The grey region shows the `cosmic variance',
the inherent statistical uncertainty in the measurements arising from
the simple fact that we can only ever measure one sky.}\label{fig1}

\end{figure}

One of the key applications of the WMAP data is to constrain
cosmological models. A `standard' model has established itself over
the last few decades, consistent with observations from galactic
scales to the largest scales observable, in which the universe is
spatially flat, and homogeneous and isotropic on large scales, and
comprises radiation, normal matter (electrons, baryons, neutrinos),
non-baryonic cold, dark matter, and dark energy.

In addition to the matter constituents, WMAP also tests several
important predictions of the inflationary scenario. Inflation predicts
that the universe is spatially flat and that fluctuations in radiation
and matter energy density are Gaussian with a nearly scale invariant
spectrum, $n_{s}\approx 1$.

WMAP is a critical test of these models, and finds them in good
agreement with the data. Under the assumption of flatness the CMB can
constrain a range of parameters on its own: the Hubble constant,
$H_{0}=100 \ h$ km/s/Mpc, is found to be $h=0.72\pm0.05$ (all error
bars are at the 68\% level), the universe is found to have an age of
13.4 $\pm$ 0.03 Gyr. For a measure of the dark matter density today,
as a fraction of the critical density (to give flat spatial curvature)
$\Omega_{m}$, WMAP finds $\Omega_{m}h^{2}=0.14\pm0.02$, and similarly
for the fractional baryon density $\Omega_{b}h^{2}=0.024\pm0.001$,
this latter one in good agreement with constraints from
nucleosynthesis. The optical depth to the decoupling surface, $\tau$,
determined by the history of recombination and re-ionization, is also
constrained although it is highly degenerate with the spectral tilt,
$n_{s}$. WMAP has made the first measurements of CMB polarization that
can be used as an independent measurement of $\tau$ and seems to give
the strongest evidence yet for an epoch of re-ionization. With TT and
TE data combined WMAP finds, $\tau=0.166^{+0.076}_{-0.071}$ and
$n_{s}=0.99\pm0.04$. The value of $\tau$ signals that re-ionization
occurred earlier than previously expected, at around a redshift of
17$\pm 5$.  Early re-ionization implies that structure was forming at
these redshifts providing evidence against the presence of significant
warm dark matter which would suppress structure formation until much
later times.

A spectral index close to unity is one finding that is consistent with
inflation. In addition to this WMAP also finds that the fluctuations
are entirely consistent with Gaussianity, and have placed the tightest
constraints yet on the level of non-Gaussianity within the primordial
spectrum. Testing the inflationary prediction of flatness is made
difficult by the presence of a geometrical degeneracy between the
fractional energy densities of spatial curvature and dark energy. A
determination of the spatial curvature and dark energy contributions
can only be obtained by breaking this degeneracy through the inclusion
of independent data sets such as the HST Key Project measurement of
$h=0.72\pm 0.05$ [5]. The data then shows a strong preference
for flatness ($\Omega_{tot}=1$) finding $\Omega_{tot}=1.02\pm0.02$. In
combination with complementary data sets, the WMAP data implies that
the universe today is made up of 73\% dark energy, 22\% dark matter
and 4.4\% baryons.

The standard models described above employ the smallest number of
parameters to fit the data, however the CMB in combination with
external data sets can be used to probe beyond these to more exotic
models. One good example of this is the placing of constraints on the
equation of state of dark energy, $w$; the additional inclusion of
supernovae observations indicates $w<-0.78$, and is entirely
consistent with the presence of a cosmological constant $\Lambda$
which has $w=-1$. WMAP, in combination with the 2dF galaxy [6]
and Lyman $\alpha$ [7] power spectra, tends to favor a varying
spectral tilt i.e. d$n_{s}$/d$\ln k \neq 0$. This variation is a
prediction of inflation but further analysis and data will help to
ascertain if the effect really is arising from subtleties of the
primordial spectrum.

WMAP continues to collect data and its planned operation is for at
least 4 years. It is hoped that this will lead to even better
understanding of systematics, better resolution at smaller scales and
improved measurement of the polarization. We are looking forward to an
exciting era in cosmology promising the elucidation of the matter
content and ionization history of the universe as well as a clearer
understanding of the inflationary epoch.

{\bf References:}

[1] Bennett, C. et al., accepted by ApJ, astro-ph/0302208.

[2] Spergel, D. et al., 
accepted by ApJ, astro-ph/0302209.

[3] Bennett, C. et al.,  ApJ, {\bf 396},  (1992) L7.

[4] Kovac, J. et al.,  Nature {\bf 420} (2002) 772.

[5]  Freedman, W. L. et al., ApJ {\bf 553} (2001) 47.

[6] Percival, W. J. et al., MNRAS {\bf 327} (2001) 1297.

[7] Croft, R. A. C. et al., 
ApJ {\bf581} (2002) 20; Gnedin N. Y. et al., 
MNRAS {\bf 334} (2002) 107.

\vfill\eject
\section*{\centerline {
Short-Range Searches for Non-Newtonian Gravity}}
\addcontentsline{toc}{subsubsection}{\it 
Short-Range Searches for Non-Newtonian Gravity, by Michael Varney and
Joshua Long}
\begin{center}
Michael C. M. Varney, Univ. Colorado, Boulder \\ 
Joshua C. Long, Los Alamos Neutron Science Center 
Collaboration 
\htmladdnormallink{Michael.Varney@colorado.edu}
{Michael.Varney@colorado.edu},
\htmladdnormallink{josh.long@lanl.gov}
{josh.long@lanl.gov}
\end{center}

In 1667, Isaac Newton proposed his famous universal law of gravity: 
\begin{equation}
F = -\frac{G m_{1}m_{2}}{r^{2}},
\end{equation}
where $F$ is the force between test masses $m_{1}$ and $m_{2}$,
$r$ is their separation, and $G$ is the gravitational constant.  
But how universal is this law?  Tests of Newtonian gravity and
searches for new macroscopic forces have covered length scales from 
light-years to nanometers, and it has been found that new forces of 
gravitational strength can be excluded for ranges from 200 microns to
nearly a light-year [1,2].  Limits on new
forces become poor very rapidly below 100 microns.

During the last 5-7 years there has been a surge of interest in
testing Newtonian gravity at sub-millimeter scales, based on many
specific theoretical predictions of modifications to gravity in this
regime.  Most notable are possible signatures of ``large'' extra
dimensions which could modify gravity directly below the millimeter
range [3].  Additional sub-millimeter effects of gravitational strength
and substantially stronger are predicted to arise as a consequence of
new particles propagating in the extra dimensions [4]. In 
older string theory-inspired models with low-energy supersymmetry 
breaking, massive scalar particles including moduli and dilatons 
are predicted to mediate new short-range forces [5]
Other predictions arise in models attempting to explain the observed 
smallness of the cosmological constant [6].

These developments have motivated a variety of novel table-top
experiments, and there has been substantial progress in improving the 
limits on non-Newtonian effects over the past three years.
Experimental results are usually parameterized with the Yukawa
interaction.  The potential due to gravity plus an additional Yukawa
force is given by: 
\begin{equation}
V = -\frac{G m_{1}m_{2}}{r}[1+\alpha\exp(-r/\lambda)],
\label{eq:yukawa}
\end{equation}
where $\alpha$ is the strength of a possible new interaction 
relative to standard gravity and $\lambda$ is the range.  The
experiments cover a range of about seven orders of magnitude, from a few 
nanometers to a few centimeters, use a variety
of techniques and confront different backgrounds.  The authors thought
it would be useful to attempt a short summary of the progress and
prospects of tests for new short-range forces, covering a broad range
of the recent small experiments.  For more detailed
reviews of various subsets of the experiments, see
Refs. [7, 13, 20].

{\it Low Frequency Experiments}: The sensitive and linear response of
the torsion balance has made it the instrument of choice for laboratory
gravity measurements.  Low frequency operation makes for low thermal
noise,  but presents challenges for vibration isolation which is also
important for attaining small test mass separations.

1. Eric Adelberger at the University of Washington, along with Blayne
Heckel and graduate students Dan Kapner and C. D. Hoyle (now at
U. Trento, Italy), operated an exquisitely
designed ``missing mass'' torsional pendulum experiment.  This
experiment provides the limits $\alpha = 10$ at $\lambda = 100$ 
microns to $\alpha = 10^{-2}$ at $\lambda = 3$~mm, which are the best 
currently published in that range [2].  
The force-sensitive pendulum mass consisted of a 1~mm thick aluminum annulus 
with an array of 10 equally spaced holes. The source mass was a stack
of two copper disks,
each a few millimeters thick, with similar arrays of holes and which
rotated approximately once every two hours.  The source mass torqued
the pendulum 10 times per revolution allowing easy discrimination from 
vibrations associated with the source drive. The source mass and
detector pendulum were separated by a 20 micron thick beryllium copper 
stretched membrane, with a total test mass separation of 197 microns. 

Current efforts are underway to allow increased sensitivity and closer 
test mass separation [2,7]. To
facilitate this, a new design utilizing a higher density ring and
attractor (copper and molybdenum, respectively) with 22 fold symmetry 
has been constructed.  Noise has been improved by a factor of six over 
the previous experiment, and a passive ``bounce-mode'' damper has been
employed, allowing test mass separation to be reduced by a factor of
2. This new experiment should be able to probe gravitational strength
forces down to 60 microns.

2. The best limits in the range from 3~mm ($\alpha = 10^{-2}$) 
to 3~cm ($\alpha = 10^{-4}$) still derive from
the null-geometry torsion balance experiment of R. Newman and
colleagues at the University of California at Irvine [8].
Torque sensitivity was limited by tilt errors and other important
systematics included magnetic and seismic effects.  Recently, the 
Irvine group, in collaboration with P. Boynton of the University of 
Washington, has constructed a torsion pendulum designed to operate at
$\sim$ 2~K in a seismically quiet underground site at
Hanford [9].  The low temperature allows operation of the 
pendulum in the tilt-insensitive frequency mode [10],
and for superconducting shielding to reduce magnetic
backgrounds.  The experiment, when operated as a test of the inverse
square law (a precision measurement of $G$ is planned first), 
will be most sensitive to new forces at a range of about
15~cm and in the thermal noise limit is expected to achieve a sensitivity at 
least two orders of magnitude greater than previous experiments at
that range.

3. Ho Jung Paik along with M. Vol Moody and colleagues at the
University of Maryland are prototyping the cryogenic ISLES (Inverse
Square Law Experiment in Space) project, to be conducted on the
International Space Station [11].  In this experiment, two
superconducting magnetically levitated niobium test disks are
suspended 100 microns on either side of a radially mounted tantalum
source disk with thin superconducting shields in between to suppress
electromagnetic couplings.  The source is nominally 140~mm in diameter
and 2~mm thick.  The planar geometry permits concentration of as much
mass as possible at the short range to be explored, and represents a 
nearly null geometry for Newtonian background forces.  The differential
acceleration of the test masses is measured as the
source is driven magnetically into small oscillations along its axis
at about 0.2~Hz, using a SQUID readout similar to the instrumentation 
developed for the Maryland superconducting gravity gradiometer.  The 
projected sensitivity, ranging from $\alpha = 10$ at $\lambda = 5$
microns to $\alpha = 10^{-5}$ at 200 microns, is up to 7 orders of 
magnitude stronger than current limits in this range.  
This sensitivity is also sufficient to further constrain (or
possibly detect) the axion, a light pseudoscalar proposed as a 
solution to the strong CP problem of QCD.  

The low-g environment of the ISS is essential for the magnetic
suspension of the test masses, which in turn provides the extremely
soft vibration isolation partly responsible for the impressive
sensitivity.  A further improvement of up to three orders of magnitude
might be possible in a free-flyer version of the experiment.
Mechanical suspension will be used in the ground-based prototype
currently under construction, reducing the projected sensitivity by up
to two orders of magnitude relative to the ISS version.

{\it High-Frequency Experiments}: Recent experiments designed to
operate in the range of a few hundred to a thousand Hz show promise to
operate at the thermal noise limit and to attain extremely small
test mass separations.

4. Early this year, the authors and their advisor John Price at the
University of Colorado published results from their high frequency
experiment, giving the current best limits between 20 and 
100 microns [12].  Forces greater 
than $10^{4}$ times gravity at 20 microns and greater than 10 times gravity 
at 100 microns were excluded. 

The apparatus uses a nominally null planar geometry and operates 
at room temperature under a vacuum of $\sim 10^{-7}$ torr to reduce
acoustic backgrounds.  A 35~mm $\times$ 7~mm $\times$ 0.305~mm 
tungsten ``diving board'' cantilever is driven at the resonant
frequency ($\sim$ 1~kHz) of a high-Q 
($\sim$ 25500) compound torsional tungsten detector oscillator.  The test
masses are separated by a 0.06~mm thick gold plated sapphire shield to
suppress electrostatic and acoustic backgrounds.  The amplitude of the 
detector mass is read via a capacitive transducer.  Measurements were 
taken at a test mass separation of 108 microns with a
test mass overlap area of about 58~mm$^{2}$, and were found to be
thermal-noise limited. 

Recently, the system has been redesigned to make use of a 10 micron
thick gold plated copper stretched membrane as an
electrostatic shield.  This will allow for mass separations of about 50 
microns, with the goal of improving the limits at that range by at 
least an order of magnitude [13].  Possible plans for the
future include a 4.2~K version of the experiment, higher Q tungsten 
detectors, and improved flatness of the test masses.

5. Aharon Kapitulnik's group at Stanford recently published results
from their high frequency cryogenic experiment [14].  This 
experiment utilized a silicon nitride microcantilever with a 50 micron 
gold cube mounted on its free end as a high Q detector. A planar
source mass, consisting of alternating gold and silicon strips, was
driven in the direction transverse to the cantilever mode of the
detector.  The alternating strip design permitted the source to be
driven at a frequency below the cantilever resonance, reducing the
burden on vibration isolation.  Detector mass amplitude was read via
an optical fiber. Operating at 10~K, the system obtained 
a sensitivity of around $8.9 \times 10^{-17}$~N at a 25 
micron source/test mass gap.  

A spurious signal, most likely electrostatic in nature, limited the
sensitivity of their apparatus.  As the phase of this signal was not
consistent with a new non-Newtonian force, the results were used to
derive limits $\alpha = 10^{9}$ at $\lambda = 3$ microns to $\alpha =
10^{5}$ at $\lambda = 10$ microns, the most sensitive in that range.
The experiment was designed to be a ``null'' experiment limited by
thermal noise with a predicted sensitivity about an order of magnitude
below that of the published results.  Re-design of the source mass to
remedy the background is in progress.

6. S. Schiller, L. Haiberger and colleagues at the University of
Dusseldorf have constructed a room temperature prototype of a high
frequency experiment of great potential sensitivity.  The detector
mass consists of a high purity silicon wafer similar in shape and size
to that of the Colorado experiment.  The use of silicon is expected to
lead to very high Qs, especially in a planned cryogenic version of the
experiment. The source mass consists of a variable-density rotor with
a harmonic corresponding to the resonance frequency ($\sim$ 5~kHz) of
the detector mass.  The source and detector mass are mounted
vertically and are separated by a conducting plate serving as an
electrostatic shield.  Detector oscillations are read out via an
optical system.

Recent measurements yielded signals about 10 times the detector
thermal noise level, corresponding to a sensitivity ranging from about
$\alpha = 10^{9}$ at $\lambda = 60$ microns to $\alpha = 5000$ at 
$\lambda = 5$~mm~ [15].  Improvements to the vibration 
isolation and thermal stability are under way.  The cryogenic
experiment is anticipated to reach gravitational sensitivity at 50 microns.

{\it Casimir Force Measurements}: For test mass separations below
about 10 microns, the Casimir effect, a force which arises between 
conductors due to zero-point fluctuations of the electromagnetic
field, becomes significant.  While of experimental interest in its own 
right, the Casimir force presents a dominant background to
experimental searches for new effects at very short ranges, and must 
either be precisely characterized or suppressed (or both) if these 
experiments are to attain greater sensitivity. 

7. Umar Mohideen and his group at the University of California at
Riverside have continued to refine their precision measurements of the 
Casimir force using AFM techniques.  In a recent experiment, a 200
micron diameter gold-plated sphere was attached to an AFM cantilever
and suspended above a gold-plated sapphire disk [18]
Deflection of the cantilever under the influence of the plate-sphere
force was monitored optically, for plate-sphere separations ranging
from 0 to 400~nm.  While the absolute force errors were slightly 
larger than for previous measurements, the use of denser test masses 
led to the strongest constraints yet attained in the range considered 
($\alpha = 10^{19}$ at $\lambda = 20$~nm to $\alpha = 10^{14}$ at
100~nm) when the results were compared with theory.

More recently this group has used similar apparatus to measure the
lateral Casimir force between a gold-plated sphere and a sinusoidally 
corrugated gold surface [19].  Good agreement with theory 
was obtained, though the constraints on new effects are less
sensitive.  Work continues on third-generation Casimir force
experiments with improved precision.

8. V. Mostepanenko's collaboration with Riverside is part of a more
general program in which he and his and colleagues have continued to
derive constraints on new physics by comparing Casimir force
measurements with the most sophisticated theoretical models.  Much of
their recent work is summarized in a comprehensive
review [20].  They have also collaborated with
E. Fischbach's group [see below] in deriving constraints in the very
short range just above 1~nm~ [21].  In this regime, the most
sensitive limits ($\alpha \approx 10^{25}$) come from a recent Casimir
force measurement by Ederth using crossed cylinders [22].

9. A group from the INFN, Padova and Pavia, and the University of
Padova run an experiment that is an interesting hybrid of high
frequency cantilever and direct Casimir measurement. The
force-sensitive detector is a 47 micron thick, 2~cm long silicon
cantilever with a 50~nm thick chromium plating. It is driven
electrically at its lowest order resonance mode of 138~Hz.  The source
mass is a 5~mm thick chromium-plated silicon block which is brought to
within less than a micron of the detector surface.  By optically
monitoring the detector resonant frequency shifts induced by the
static external force gradient, the group was able to obtain a
measurement of the Casimir force to a precision of 15\%, the first
measurement of this effect for the parallel-plate geometry [16].
Limits on new effects derived from this experiment are not quite
competitive with the most stringent constraints in the relevant range
near 1 micron, but an optimized version of this experiment under
construction is expected to improve these limits by at least an order
of magnitude [17].

10. Ephraim Fischbach and colleagues at Purdue University including
Dennis Krause (also at Wabash College) are pursuing experimental and
theoretical programs to control the Casimir background in short-range
experiments.  Sub-micron measurements of the differences in forces
between different isotopes of the same element are underway.  These
are expected be sensitive to new short-range effects as the Casimir
force should be dominated by the electronic properties of the test
masses and essentially independent of isotope (iso-electronic effect).

In a recent theoretical study [23], this group has
quantified the isotopic dependence of the Casimir force, and estimated
the fractional difference in Casimir forces between two isotopes of
the same element to be on the order of $10^{-4}$.  This is roughly two
orders of magnitude below the resolution of recent Casimir force
measurements, lending confidence to the prospect that differences
observed in iso-electronic experiments will be due to other effects.

An initial experiment (designed primarily to investigate gross
systematics and sample fabrication) used an AFM to measure the forces
on a silicon nitride cantilever suspended a few nm above a surface
consisting of alternating regions of gold and copper [24]
(These metals have very similar electronic properties but different
densities.)  Results from these measurements were used to set limits
of $\alpha \approx 10^{27}$ in the range $\lambda$ = 1-2~nm, slightly
more sensitive than the previous limits in that range.

More recently this group has reported the first precise measurement of
the Casimir Force between dissimilar metals [25].  This
experiment used a more sensitive microelectromechanical apparatus, in
which a 600 micron diameter gold-plated sphere was suspended above one
side of a .25~mm$^{2}$ copper-coated torsional plate.  Forces were
measured as a function of plate-sphere separations from about 200 to
1200~nm, by observing both the static deflection of the plate and the
change in its resonant frequency as it was driven into small
oscillations.  Comparison of these measurements with a detailed
theoretical model leads to the limit of $\alpha \approx 10^{13}$ at
$\lambda$ = 200~nm, about a factor of 4 improvement over the previous
limit at that range.  These results are based on an observed
systematic difference between theory and experiment. The group
suspects this is based on the imprecise characterization of the
optical properties of the metals, and emphasizes the need for better
measurements of these properties and better theoretical understanding
of the Casimir force for non-ideal objects.  Plans to improve the
limits by comparing the force on the sphere to two isotopes of the
same element are also underway.

In summary, the authors are aware of at least 10 active programs
pursuing short range experiments with implications for 
non-Newtonian gravity.  Over the next few years, prospects are good 
for improving the limits on new effects by several orders of 
magnitude in the range from nanometers to centimeters.  
The authors wish to thank all researchers who replied to requests for
the latest news.

\parskip=3pt
{\bf References:}

[1] Fischbach E., Talmadge C., {\it The Search for
  Non-Newtonian Gravity} (New York, Springer-Verlag, 1999) 62.

[2] Hoyle C. D., Schmidt U., Heckel B. R., Adelberger E. G., Gundlach
  J. H., Kapner D. J., Swanson H. E., Phys. Rev. Lett. 86 (2001) 1418;
  Adelberger E. G., arXiv hep-ex/0202008.

[3] Arkani-Hamed N., Dimopoulos S., Dvali G.,
  Phys. Lett. B 429 (1998) 263; Randall L., Sundrum R.,
  Phys. Rev. Lett. 83 (1999) 4690.

[4] Arkani-Hamed N., Dimopoulos S., Phys. Rev. D 65
  (2002) 052003.

[5] Dimopoulos S., Giudice G., Phys. Lett. B 379 (1996)
  105; Taylor T. R., Veneziano G., Phys. Lett. B 213 (1988) 450; 
  Kaplan D. B., Wise M. B., J. High Energy Phys. 0008 (2000) 037.  

[6] Sundrum R., J. High Energy Phys. 7 (1999) 1; Schmidhuber
  C., arXiv:hep-th/0207203.

[7] Adelberger E. G., Heckel B. R., Nelson
  A. E., arXiv hep-ph/0307284. 

[8] Hoskins J. H., Newman R. D., Spero R., Schultz J.,
  Phys. Rev. D 32 (1985) 3084. 

[9] Newman R., Class. Quantum Grav. 18 (2001) 2407.

[10] Boynton P. E., Class. Quantum Grav. 17 (2000) 2319.

[11] Paik H. J., Moody M. V., Strayer D. M., Proceedings
  of the 2002 NASA/JPL Workshop on Fundamental Physics in Space (Dana
  Point, California, May 2002).

[12] Long J. C., Chan H. W., Churnside A. B., Gulbis
  E. A., Varney M. C. M., Price J. C., Nature 421, (2003) 922.

[13]  Long J. C., Price J. C., C. R. Physique 4
  (2003) 337.

[14] Chiaverini J., Smullin S. J., Geraci A. A., Weld
  D. M., Kapitulnik A., Phys. Rev. Lett. 90 (2003) 151101.

[15] Haiberger L., Weingram M., Wenz H., Schiller S.,
  ``An experiment to detect gravity at sub-mm scale with high Q
  mechanical oscillators,'' presented at the Tenth Marcel Grossmann
  Meeting on General Relativity, Rio de Janeiro, 2003.

[16]  Bressi G., Carugno G., Onofrio R., Ruoso G.,
  Phys. Rev. Lett. 88 (2002) 041804.

[17] Bressi G., Carugno G., Galvani A., Onofrio R.,
  Ruoso G., Veronese F., Class. Quantum Grav. 18 (2001)

[18] Harris B. W., Chen F., Mohideen U., Phys. Rev. A
  62 (2000) 052109

[19] Chen F., Mohideen U., Klimchitskaya G. L.,
  Mostepanenko V. M., Phys. Rev. A 66 (2002) 032113

[20] Bordag M., Mohideen U., Mostepanenko V. M.,
  Phys. Rept. 353 (2001) 1. 

[21]  Fischbach E., Krause D. E., Mostepanenko V. M.,
  Novello M., Phys. Rev. D 64 (2001) 075010. 

[22]  Ederth T., Phys. Rev. A 62 (2000) 062104

[23]  Krause D. E., Fischbach E., Phys. Rev. Lett. 89 (2002) 190406

[24]  Fischbach E. et al., Class. Quantum Grav. 18 (2001) 2427

[25] Decca R. S., Lopez D., Fischbach E., Krause D. E.,
  Phys. Rev. Lett. 91 (2003) 050402

\parskip=10pt

\vfill\eject

\section*{\centerline {
Xth Brazilian School of cosmology and gravitation}}
\addtocontents{toc}{\protect\medskip}
\addtocontents{toc}{\bf Conference reports:}
\addcontentsline{toc}{subsubsection}{\it  
Xth Brazilian School of cosmology and gravitation, by Mario Novello}
\begin{center}
Mario Novello, CBPF, Rio de Janeiro
\htmladdnormallink{novello@cbpf.br}
{mailto:novello@cbpf.br}
\end{center}

In 2002, during the Xth Brazilian School of Cosmology and Gravitation
(BSCG), we celebrated the 25th anniversary of the School, which has
been organized since 1977 by the Group of Cosmology and Gravitation of
the CBPF.  To commemorate this unique moment, a web page is being
launched with all the 93 lectures and seminars of the first nine
schools ( the proceedings of the Xth BSCG will be published this year
by AIP): more than 4500 pages in pdf format, provided by many of the
most important scientists in Cosmology, Gravitation, Astrophysics, and
Field Theory.  It is an important contribution for students and
researchers in these areas, which shows the historical evolution of
physics in the last 25 years.  This material can be accessed through
the page of the Group of Cosmology and Gravitation of the CBPF:

\htmladdnormallink
{www.cosmologia.cbpf.br}
{www.cosmologia.cbpf.br}

\vfill\eject

\section*{\centerline {
Sixth East Coast Gravity Meeting}}
\addcontentsline{toc}{subsubsection}{\it  
Sixth East Coast Gravity Meeting}
\begin{center}
David Fiske, University of Maryland
\htmladdnormallink{drfiske@physics.umd.edu}
{drfiske@physics.umd.edu}
\end{center}

The Sixth East Coast Gravity Meeting was held at the University of
Maryland, College Park on March 29 and March 30, 2003.  Thirty-five
contributed talks on a wide range of gravity-related topics filled
a one and one-half day program, attended by over fifty people
from as far north as Maine, as far south as North Carolina, and as far
west as California.
The meeting was divided into two sessions dedicated to numerical
relativity, one session covering classical general relativity and gravitational
waves, one session dealing with compact objects, and three sessions touching
on topics from the broad category of general relativity and beyond.

As has been the tradition at the East Coast Meetings, a large number
of students contributed talks to this year's meeting.  In addition to
the ten graduate students (from Brown, Maryland, Penn State, and
Syracuse) who presented, two undergraduates from Bowdoin College
contributed excellent talks.  David Mattingly from the University of
Maryland was recognized with the 
\htmladdnormallink{Best Student
Presentation Award}
{http://gravity.phys.psu.edu/~{}tggweb/regulations/ggr-best-talk.html},
sponsored by the APS Topical Group on Gravitation.  The judges also
recognized Monica Skoge, an undergraduate from Bowdoin College, with
an Honorable Mention in this year's competition.

Although there are no formal proceedings for the meeting, the program
with abstracts remains online at the conference webpage
\htmladdnormallink{www.physics.umd.edu/grt/ecgm}
{http://www.physics.umd.edu/grt/ecgm/}.
We have encouraged speakers to provide relevant references
to archived papers, and have added active links to those papers so that
interested readers can easily follow up on the content of the talks.
Speakers still wishing to add such references should send them to 
\htmladdnormallink{ecgm@physics.umd.edu}{mailto:ecgm@physics.umd.edu}.

The Seventh East Coast Gravity Meeting will be held next year at
Bowdoin College in Brunswick, Maine.  A link to the conference
webpage, when it is available, will appear on the East Coast Gravity
Meeting coordinating committee's webpage
\htmladdnormallink{http://physics.syr.edu/\~{}marolf/ECGM.html}
{http://physics.syr.edu/~marolf/ECGM.html}.  In addition, the
coordinating committee is accepting volunteers to host future
meetings.  Information about the history of the meeting and about
volunteering are also available on the committee's webpage.

This year's organizers would like to thank the University of Maryland 
Department of Physics for contributing financial support.

\vfill\eject

\section*{\centerline {
5th Edoardo Amaldi meeting}}
\addcontentsline{toc}{subsubsection}{\it  
5th Edoardo Amaldi meeting, by Alain Brillet}
\begin{center}
Alain Brillet, Virgo Project
\htmladdnormallink{brillet@obs-nice.fr}
{brillet@obs-nice.fr}
\end{center}

The 5th Edoardo Amaldi conference on Gravitational Waves took place in
 Tirrenia (Tuscany) on July 6th to 11th. The facilities and the
 organization were excellent.  Tours to the site of Virgo, only 15 km
 away, were organized everyday.  Mainly dedicated to experimental
 activities in the field of gravitational wave detection, the
 conference attracted more than 250 participants, mostly
 experimentalists from all the ongoing detector projects.  

After a
 session of overviews, six sessions were dedicated to the detectors
 (status, advanced techniques, fundamental noise sources, and future
 detectors), one to theory and sources, and one to data analysis.

 Since the last conference two years ago in Australia, the general
 progress is remarkable: the LIGO large interferometric detectors (2
 in Hanford, Wa, and one in Livingston, La) have been all completed
 and are getting close to the expected sensitivity at high frequency,
 the construction of Virgo, in Tuscany, has included a successful
 pre-commissioning period and is now completed, the TAMA and GEO
 prototype/detectors are also running. If the sensitivity of the
 interferometers still remains to be improved, particularly in the low
 frequency range, the good news are that they all demonstrate a good
 duty cycle: when the site activity stops for an ``engineering run" or
 a ``science run", they do run unattended for hours to weeks. This is
 quite remarkable, given the complexity and the delicacy of these
 instruments. The analysis of the first data from LIGO, TAMA, and GEO
 detectors allows the determination of upper limits on the amplitude
 and event rate for various possible GW sources. Although these limits
 are not yet quantitatively interesting, they demonstrate the ability
 to acquire and store the data, and to filter them with appropriate
 templates. These studies have also the merit to show that the final
 sensitivity (after data filtering) is strongly affected by the lack
 of stationarity of the noise. This has a consequence on the detailed
 design of the interferometers, that the Tama group has started to
 take into account: it is important to improve the noise stationarity,
 as well as the noise spectral density.  

GEO plays the double role of
 a detector, and of an advanced prototype: after a ``science run" in
 coincidence with LIGO, it will implemented as a double recycling
 interferometer, using advanced technologies in the last suspension
 stages: monolithic fiber suspensions, controlled damping of the
 violin modes, thermal control of the curvature radius, …The
 Australian AIGO group also develops advanced instrumentation in
 collaboration with LIGO and with Virgo.  

``Advanced techniques" and
 ``Noise sources" were mainly focused on two points: thermal noise and
 quantum noise. Thermal noise issues are the most immediate concern:
 since the previous Amaldi meeting in Perth, it was found that the
 mirror's thermal noise, due to the mechanical losses of the
 reflective coatings, could become relevant, particularly in advanced
 uncooled detectors like the Advanced LIGO. It may limit the
 sensitivity if the planned sapphire substrates are not yet available
 and have to be replaced by silica. If the material of the coating
 layers cannot be sufficiently improved, it remains the possibility of
 using non-Gaussian beams, which should provide a lower phase noise
 level, for the same mirror fluctuations. This thermal noise would
 anyway become negligible in a cryogenic detector, were quantum noise
 should be the main limitation. Experimental and theoretical studies
 were presented on the quantum noise problems. ``Quantum locking", a
 new idea, which consists in measuring the radiation pressure noise on
 each mirror and providing active feedback, does not require the use
 of squeezing or QND techniques. The first large cryogenic prototype,
 CLIO, is being built, underground, in the Kamioka mountain.

Except for Nautilus and Explorer the acoustic detectors have not been
running much in the last years, but good progress has been made in the
noise and the bandwidth of the transducers, which will keep them
competitive for some more time with the wideband interferometers.  A
promise for the future is the development of spherical cryogenic
detectors: three of them may be built soon, in Leiden (mini-Grail), in
Sao Paulo (Mario Schenberg) and in Rome (Sfera), involving two new
groups and new countries in the field.

One full session was dedicated to LISA, and to the SMART-2 test
flight, foreseen in 2006-2007. The spacecraft will carry two similar
but different payloads, one built by NASA and one by ESA. They will
both test the functionality and the sensitivity of the
interferometers, of the ``gravity reference sensors", and of
micro-thrusters to be used later in LISA. The progress on these
critical issues is impressive. The collaboration-competition
compromise is not yet totally clear.

The session on theory and GW sources was mainly focused on the
description of new possible sources to be detected by LISA and by
LIGO-Virgo and the corresponding advanced detectors, on the
possibility to use LISA signals and their redundancy for testing
alternative theories, and on the progress in numerical
relativity. Recent achievements, and new investments in computing
facilities, should allow soon for an interesting comparison between
''exact" models and real coalescence signals.

The data analysis session became very large, with (too) many (too)
short talks, and more than 20 posters. At the next issue of the Amaldi
conference ( Summer 2005, in Japan), it may be wise to focus on the
results of the working detectors (upper limits , noise studies,…)
and to push other papers towards the GWDAW (Gravitational Wave Data
Analysis Workshop) that the community organizes each year in December.
The session on future detectors proposed a full description of the
Advanced LIGO, and considerations about Japanese and European projects
for cryogenic detectors, improved SQUID's for acoustic detectors, and
even a ``post-LISA" proposal. Funds and manpower may be limited, but
the research field is young, and there is no lack of new ideas!

\vfill\eject

\section*{\centerline {
Pacific Coast Gravity Meeting}}
\addcontentsline{toc}{subsubsection}{\it  
Pacific Coast Gravity Meeting, by Charles Torre}
\begin{center}
Charles Torre, Utah State
\htmladdnormallink{torre@cc.usu.edu}
{torre@cc.usu.edu}
\end{center}

The $19^{th}$ Pacific Coast Gravity Meeting (PCGM) was held at the
University of Utah, February 28 -- March 1. The conference was held in
conjunction with a celebration of Richard Price's sixtieth birthday --
the Pricefest -- which added an additional dimension to the
meeting. In particular, the meeting benefited from overlapping with
some of the Pricefest events, which included a party at Richard
Price's house on Friday night as well as a solemn, respectful tribute
dinner for Richard on Saturday night. See John Whelan's article
elsewhere in this newsletter for details on the Pricefest.

The format of this meeting was the same as usual for this small,
informal regional meeting: a wide variety of 15 minute talks from over
40 participants. The conference was very smoothly run by postdoctoral
``volunteers'' Chris Beetle and Lior Burko. As always, this conference
tries to encourage student participation, and there were twelve
student talks at this year's meeting. Nothing is quite so encouraging
as money, and a cash prize named after Jocelyn Bell is awarded for the
best student talk. For the first time (as far as I know), this prize
was funded by the American Physical Society Topical Group on
Gravitation. This year's Bell prize was awarded to Henriette Elvang
(University of California, Santa Barbara) for her talk on ``Bubbles
and Black Holes''.  Last I heard, the $20^{th}$ Pacific Coast Gravity
Meeting will be held in late winter 2004 at Caltech.

Several (overlapping) themes were represented at this year's
conference. Here is an attempt to identify those themes and the
speakers that go with them.  Of course, several talks could be put
into more than one category.

{\it The Gravitational Two-Body Problem} with
talks by John Baker, Chris Beetle, Ben Bromley, Lior Burko,
Rachel Costello, Peter Diener, Carlos Lousto, Mark Miller,
Richard Price, Charles Torre.

{\it LIGO/LISA} with talks by Patrick Brady, Teviet Creighton,
 Jonathan Gair, Louis Rubbo, Linqing Wen, John Whelan
 
{\it Experimental Gravity -- but not LIGO or LISA} with presentations
by Eric Berg, Liam Cross, Daniel Kapner, Frank Marcoline, 
Jason Steffan, Clifford Will.  

{\it Gravitational Collapse and Numerical Relativity -- but not
the 2-body problem} with talks by 
Jay Call, R. Steven Millward, Frans Pretorius, Mark Scheel, Richard O'Shaughnessy.

{\it String Theory,  Quantum Gravity, and then some} with
presentations by 
 Steve Carlip 
, Yujun Chen 
, Tevian Dray
, Henriette Elvang 
, Gary Horowitz
, Karel Kucha\v r
, Alok Laddha 
, Chad Middleton 
, Jorge Pullin.

{\it Mathematical Properties of the Einstein Equations and Their
Solutions} with talks by 
Beverly Berger,
Simonetta Frittelli,
Jim Isenberg,
Lee Lindblom.

{\it Other Topics in Field Theory and Gravitation}
with presentations by  Randy Dumse and William Pezzaglia

\vfill\eject

\section*{\centerline {Astrophysics of Gravitational Wave Sources
Workshop}}
\addcontentsline{toc}{subsubsection}{\it  
Astrophysics of Gravitational Wave Sources
Workshop, by Joan Centrella}
\begin{center}
Joan Centrella, NASA-Goddard
\htmladdnormallink{Joan.Centrella@nasa.gov}
{Joan.Centrella@nasa.gov}
\end{center}

On April 24 - 26, 2003, a group of researchers from around the world
gathered at the University of Maryland's Inn and Conference Center for
the workshop ``The Astrophysics of Gravitational Wave Sources."  The
speakers and attendees represented a broad range of areas within
physics and astrophysics: general relativity; optical, X-ray, and
G-ray astronomy; theoretical astrophysics; gravitational wave
detection; and data analysis. The talks and discussions were
stimulating and informative, covering diverse areas of this growing
field.  In fact, the workshop provided an opportunity for many of
those present to meet and interact with each other for the first time.

The meeting began with an overview of gravitational wave sources.
This was followed by a presentation on gravitational wave detectors
and detection circa 2012, which is the time frame in which the
advanced ground-based detectors (probing high frequency sources) and
LISA (observing low frequency sources) should be operating.  Most of
the remaining of the talks focused on the astrophysics of anticipated
gravitational wave sources and the scenarios that surround them,
including collapses, binaries, and gamma-ray bursts.  Black holes -
ranging from stellar, to intermediate mass, to supermassive - figured
prominently in many presentations. Talks on data analysis and
detection, including a report on the recent S1 run of LIGO/GEO,
rounded out the program.  Electronic versions of many of the workshop
presentations can be found online at

\htmladdnormallink
{http://astrogravs.gsfc.nasa.gov/docs/agws\_workshop/presentations/}
{http://astrogravs.gsfc.nasa.gov/docs/agws\_workshop/presentations/}.

These are exciting times.  LIGO has just completed its second
scientific data-taking run (S2), and plans for advanced ground-based
detectors are in progress.  The space-based LISA is moving forward
strongly as a partnership between NASA and ESA.  Gravitational wave
astrophysics is a stimulating and fruitful area of interaction for
researchers from diverse areas of physics and astrophysics.  The
presentations at this workshop provided snapshots of this emerging
field today, and glimpses of the scientific excitement to come.

\vfill\eject

\section*{\centerline {
Gravitational interaction of compact objects}}
\addcontentsline{toc}{subsubsection}{\it  
Gravitational interaction of compact objects, by
Matt Choptuik, \'Eanna Flanagan and Luis Lehner}
\begin{center}
Matt Choptuik, UBC, \'Eanna Flanagan, Cornell and Luis Lehner, LSU
\htmladdnormallink{choptuik@physics.ubc.ca}
{mailto:choptuik@physics.ubc.ca},
\htmladdnormallink{flanagan@astro.cornell.edu}
{mailto:flanagan@astro.cornell.edu},
\htmladdnormallink{lehner@lsu.edu}
{mailto:lehner@lsu.edu}
\end{center}
\parskip=3pt

The ``Gravitational Interaction of Compact Objects'' conference
took place at the Kavli Institute of Theoretical Physics
from May 12th to May 14th, 2003. Aside from bringing together
researchers in the field to discuss the status of activities in
the area, the conference served to ``kick-off'' a two-month
workshop under the same name at KITP.
Each day of the conference was divided into sections that focused on
specific areas, which ranged from broad themes to particular issues. 

The first day consisted of a series of talks to survey the status of
different aspects of gravitational wave data astronomy.  Fred Raab
described the status of the LIGO detectors, and Joan Centrella gave an
overview of the possible sources for both space based and ground based
detectors.  Lee Lindblom reviewed aspects of accretion induced
collapse of white dwarfs and the marginal likelihood of detection of
gravitational waves from these systems with advanced interferometers.
Following these there were four review talks on the dynamics of binary
systems.  Oliver Poujade reviewed post-Newtonian computations of the
gravitational waveform, based on matching a post-Newtonian expansion
of the field equations in the near zone to a post-Minkowskian
expansion in the far zone.  Currently there are regularization
ambiguities which arise at post-3-Newtonian order, and Poujade
described ongoing attempts to resolve these ambiguities.  John Baker
described the status of the Lazarus project to simulate binary black
holes, and the present efforts and results to obtain complete
information from initial data sets describing equal mass spinning
black holes.  Luis Lehner gave an overview of different approaches to
numerical relativity and to the binary black hole problem, and Miguel
Alcubierre reviewed the status of binary black hole simulations within
the Cauchy approach.

The second day's talks were mostly devoted to the current
understanding of specific sources, together with one talk on data
analysis.  Phil Arras reviewed the generation of gravitational
radiation by neutron star Rossby modes and the role gravitational
waves can play in determining the observed/inferred properties of
neutron stars.  He argued that r-modes in low mass X-ray binaries
(LMXBs) are a possible candidate for detection with LIGO II. This
depends somewhat on unknown properties of the star's viscosity that
govern the stability of a steady state solution in which the mode is
excited.  Detectable gravitational waves from LMXBs may also be
produced via inhomogeneities in the crust.  Next, Masaru Shibata
reviewed the status of his simulations of binary neutron star mergers,
presenting simulations of both equal and unequal mass cases which
merge after a couple of orbits.  Shibata argued that useful physical
information should be obtainable from the current codes given larger
computers and more efficient computer use.  He also discussed
incorporation of more realistic physical components like improved
equations of state, neutrino cooling etc.  Christian Cardall then
surveyed the present status of supernovae simulations. He discussed
shortcomings in present codes (lower dimensionality, Newtonian
dynamics, incomplete physical processes incorporated), and reviewed
the five-years Terascale Supernova Initiative which aims to address
all these shortcomings.

The subject then switched to gravitational wave data analysis.
Patrick Brady discussed data analysis methods for four different types
of gravitational wave signals: known waveforms, unknown burst
waveforms, periodic signals, and stochastic signals.  He presented
preliminary results from the analysis by the LIGO Science
Collaboration (LSC) of LIGO's first science run, giving upper limits
on the event rates of various types of sources.

Peter Meszaros then reviewed current understanding of gamma ray bursts
(GRBs), their phenomenology, models for the central engine, and the
likelihood of detecting gravitational waves from them.  He also
described planned detectors which will increase the number of GRBs
detected and the accuracy of their sky locations.

Subsequently, Leor Barack highlighted extreme mass ratio binaries, in
which neutron stars or solar mass black holes inspiral into
supermassive black holes, as important sources for spaced based
detectors like LISA.  He discussed theoretical problems related to
obtaining accurate templates for these inspirals, and recent results
which should provide a way to compute such templates for generic
orbits around spinning black holes.  The day ended with Greg Cook's
review talk on methods for solving the constraint equations to obtain
initial data for interacting compact binaries.  He showed that new
methods based on the conformal thin sandwich approach agreed better
with post-Newtonian results than did older methods.

On the third day, talks concentrated on pressing issues related to
simulations of compact objects and highlighted several promising new
ingredients. Two talks discussed ways to improve the availability
computational power and the efficiency of its use.  Ed Seidel gave an
overview of how so-called Grid computing can provide considerable
enhanced resources for realistic simulations. Frans Pretorius surveyed
the application and promises of adaptive mesh refinement, showing
explicit examples both in Cauchy and characteristic implementations.

The next two talks turned to general relativistic hydrodynamics
simulations, and treated in depth what will be needed for accurate and
realistic results.  Mark Miller reviewed the effect on evolutions of
the choice of initial data sets used, and discussed how to estimate
the associated errors in physical observables that one extracts from
the simulation.  He illustrated these issues by showing explicit
evolutions of multiple orbits of neutron star binary systems.  Charles
Gammie discussed the incorporation of magnetic fields in simulations,
describing the implementation of the magneto-hydrodynamics equations
and presenting explicit examples of accretion tori on spinning black
hole backgrounds.

The conference then turned to new techniques that address issues which
had previously been poorly understood.  Jeffrey Winicour described the
problems associated to dealing with boundaries, and in particular the
issue of obtaining boundary conditions that make the initial/boundary
value problem well posed.  He described different efforts to achieve a
solution of this problem in special cases.  Manuel Tiglio described a
set of novel techniques which, for linear problems, guarantee
numerical stability of the implementation.  He also described how the
parameter freedom in families of formulations of Einstein's equations
could be exploited to minimize the growth of constraint violating
modes.  He presented explicit examples in test problems where the use
of this technique made a considerable improvement.

The conference served to highlight several recent advances in the
simulation of compact objects, to review current outstanding problems 
associated with these simulations, and to improve our understanding of
possible methods for overcoming the problems.  In addition, it
reviewed the context for these simulation and modeling efforts 
 to detect and analyze gravitational wave signals,
and the possible payoffs for scientific knowledge.
All of the conference talks are available online
at

\htmladdnormallink
{http:/online.kitp.ucsb.edu/online/gravity\_c03/}
{http:/online.kitp.ucsb.edu/online/gravity\_c03/}.
\parskip=10pt

\vfill\eject

\section*{\centerline {
PriceFest}}
\addcontentsline{toc}{subsubsection}{\it  
PriceFest, by John T.~Whelan}
\begin{center}
John T.~Whelan, Loyola University New Orleans
\htmladdnormallink{jtwhelan@loyno.edu}
{mailto:jtwhelan@loyno.edu}
\end{center}
\parskip=3pt

On March 2, 2003, the day after the 19th Pacific Coast Gravity
Meeting, the one-day ``PriceFest'' was held at the University of Utah
to celebrate the 60th birthday the day before of Richard Price.  The
festivities were hosted by Richard's longtime colleague Karel
Kucha\v{r} (Utah), who likened individual descriptions of Richard to
the blind men's descriptions of an elephant in the old parable.  Each
teacher, collaborator, colleague, or student could describe parts of
Richard's trunk, ears, legs and tail, with the whole picture coming
from the synthesis of the different perspectives.  This theme
continued throughout PriceFest, as each speaker presented his or her
perspective on Richard's life, work, and personality.

Kip Thorne (Caltech) began with ``The Early Richard Price'', an
account of Richard's days as a Caltech graduate student.  He painted a
picture of the years between 1965 and 1971 as a time of social and
scientific upheaval.  Quasars had just been discovered, ``black
holes'' were about to be named, and Tommy Gold had proclaimed at the
first Texas Symposium that relativists ``might actually be useful to
science''.  Kip described how Richard gained confidence in himself and
his work, from TAing a Relativity course despite never having taken
one, to deriving the power-law tails which carry away ``hair'' in
black-hole formation in the face of contemporary literature which
maintained otherwise.  He closed his tribute to an esteemed
prot\'{e}g\'{e} and family friend with the observation that ``each
generation of scientists often underestimates what the next generation
may achieve.''

Jorge Pullin (LSU) continued with ``Richard Price, Car Talk and The
American Journal of Physics'', which explored Richard's role as a
pedagogue and communicator of physical knowledge.  This consisted
mostly of a review of Richard's many contributions to the American
Journal of Physics, which are collected at
\htmladdnormallink{{http://www.phys.lsu.edu/faculty/pullin/rhpajp.pdf}}
{http://www.phys.lsu.edu/faculty/pullin/rhpajp.pdf}~.

After an interlude in which Richard's sister, who had been unable to
attend the banquet the night before, gave her belated contribution to
the roast, David Kieda (Utah) continued with ``The Hidden Side of
Richard Price'', the results of a web search which revealed Richard
Prices ranging from an 18th century congregational minister to a
private eye.

Richard's former officemate Bernard Schutz (AEI, Bernie \& the
Gravitones) provided the perspective of a contemporary with ``Catching
Flies and Journalists with Richard Price''.  A member of the class
which came to Caltech immediately after Richard's, Bernie listed as
his mentors Kip, Frank Estabrook, and Richard.  He credited Richard
with teaching him not only how to do research, but also how to deal
with the press.  At the time this included \textit{Los Angeles Times}
writer Jack Smith, who came across a Caltech chalkboard with Price's
theorem (``Everything that can be radiated, will be radiated"),
Schutz's converse (``Everything that is radiated, can be radiated") and
their synthesis (``Radiation does its own thing").

Carlos Lousto (UTB) concluded the f\^{e}te with ``The Late
Richard Price'', from the perspective of one of Richard's many younger
collaborators.  The common theme of these collaborations, ranging from
quasinormal ringing to quasistationary inspiral, seemed to be the
conversion of postdocs (including both the speaker and yours truly)
from quantum gravity to astrophysical relativity.  Carlos also gave
proper credit to the contributions of Richard's dog Sunny.

By the end of the day, the speakers had assembled a picture of a
scientist who has had a remarkable impact not only on the field but
especially on the people around him.

\parskip=10pt
\vfill\eject

\section*{\centerline {
Gravitation: a decennial perspective}}
\addcontentsline{toc}{subsubsection}{\it  
Gravitation: a decennial perspective, by Jorge Pullin}
\begin{center}
Jorge Pullin, LSU
\htmladdnormallink{pullin@lsu.edu}
{mailto:pullin@lsu.edu}
\end{center}

On June 8-12 a conference took place at Penn State, coinciding with
the 10 years anniversary of the foundation of the Center for
Gravitational Physics and Geometry. It sought to give a perspective of
events during the last decade and how they shape the future to come in
gravitational physics. About 150 people attended the conference.

Plenary lectures included a first day concentrated on aspects of
gravitational waves, ranging from the astrophysical, with Ramesh
Narayan, to experimental, with Rai Weiss, and numerical with Masaru
Shibata as speakers. The second day had talks by Sean Carroll and
Gary Horowitz on dark matter and gravitational aspects of string
theory respectively. The third day had lectures by Bernie Schutz, on
confronting gravitational wave observations with theory and Badri
Krishnan and myself on analytical aspects of numerical relativity. On
the fourth day Saul Teukolsky spoke about the next ten years in
numerical relativity, Carlo Rovelli on spin foam models and Ted
Jacobson on quantum gravity phenomenology. The last day had Roger
Penrose speaking about the mathematics of general relativity and ended
with a panel discussion. 

The afternoons had parallel sessions on quantum gravity and quantum 
field theory on curved spacetime, gravitational wave physics, quantum
geometry and its applications, numerical relativity, mathematical
relativity and  quantum cosmology.

The conference banquet had an after dinner talk by Ted Newman on
the blossoming of general relativity in the 60's and 70's. After the
talk, with Jim Hartle as master of ceremonies, several people made
impromptu remarks about the impact the PSU Center had had in their
lives and many highlighted the leadership role Abhay Ashtekar has
played as its director. 

Let us wish the PSU Center for Gravitational Physics and Geometry
ten more years as successful as the first!

\end{document}